\documentclass[aps,prl,reprint,twocolumn,floatfix,notitlepage]{revtex4-1}
\usepackage[english]{babel}
\usepackage{amsmath, amsthm, amssymb, latexsym}
\usepackage[usenames,dvipsnames]{xcolor}
\usepackage[utf8]{inputenc}
\usepackage[normalem]{ulem} 

\newcommand{\la}{\lambda}
\newcommand{\ua}{\textbf{u}_{\text{A}}}
\newcommand{\da}{\boldsymbol\nabla\cdot\textbf{u}_{\text{A}}}
\newcommand{\dap}{\boldsymbol\nabla'\cdot\textbf{u}_{\text{A}}'}
\newcommand{\uh}{\textbf{u}}
\newcommand{\dv}{\boldsymbol\nabla\cdot\textbf{u}}
\newcommand{\dvp}{\boldsymbol\nabla'\cdot\textbf{u}'}
\newcommand{\ja}{\textbf{J}_{\text{c}}}
\newcommand{\dja}{\boldsymbol\nabla\cdot\textbf{J}_{\text{c}}}
\newcommand{\djap}{\boldsymbol\nabla'\cdot\textbf{J}_{\text{c}}'}

\begin{document}

\title{Exact law for homogeneous compressible Hall magnetohydrodynamics turbulence}

\author{N. Andr\'es$^{1}$}
\author{S. Galtier$^{1,2}$}
\author{F. Sahraoui$^{1}$}
\affiliation{$^1$  LPP, CNRS, Ecole Polytechnique, UPMC Univ. Paris 06, Univ. Paris-Sud, Observatoire de Paris, Université Paris-Saclay, Sorbonne Universités, PSL Research University, F-91128 Palaiseau, France.  \\
             $^2$ Departement de Physique, Universit\'e Paris-Sud, Orsay, France.}
\date{\today}

\begin{abstract}
We derive the exact law for three-dimensional (3D) homogeneous compressible isothermal Hall magnetohydrodynamics (CHMHD) turbulence, without the assumption of isotropy. The Hall current is shown to introduce new flux and sources terms that act at the small scales (comparable or smaller than the ion skin depth) to significantly impact the turbulence dynamics. The new law provides an accurate means to estimate for the first time the energy cascade rate over a broad range of scales covering the MHD inertial range and the sub-ion dispersive range in 3D numerical simulations and {\it in situ} spacecraft observations of compressible turbulence. This work is particularly relevant to astrophysical flows in which small scale density fluctuations cannot be ignored such as the solar wind, planetary magnetospheres and the interstellar medium (ISM).
\end{abstract}

\maketitle

{\it Introduction.} Fully developed plasma turbulence theories are crucial to understand astrophysical flows that include the solar wind, the ISM and accretion flows \citep[see, e.g.][]{K1969,Sc2009,BC2013}. Due to the complexity and randomness of turbulent flows, exact mathematical results about turbulence are very few in the literature. The most important one that was derived for homogeneous incompressible magnetohydrodynamics (MHD) turbulence is the so-called 4/3 law. It relates the turbulent fluctuations at given scale $\ell$ to the rate by which energy (or other invariants of motion) is dissipated into the system  \citep{Ch1951,P1998a,P1998b}. This exact law has been widely used to quantify the energy cascade rate in solar wind turbulence~\citep{M1999,Smi2006,SV2007,Sa2008}, to predict the decay of MHD turbulence \citep{W2012} and to determine scaling exponents in measurements and numerical simulations of turbulence through the extended self-similarity (ESS) method \citep{R1993,G1997}.  However, those works are only valid for incompressible flows and limited to the MHD scales, and therefore ignore the role of density fluctuations and do not capture any small scale effect. By small scale effects we refer to the terms in the generalized Ohm' law that allow one to describe time and spatial scales that are comparable or smaller than the ion gyro-period and skin depth $d_i=c/\omega_{pi}$ (with $c$ is the speed of light and $\omega_{pi}$ is the ion plasma frequency). At those scales the MHD description breaks down, and the first-order correction that can be considered in a fluid description of plasmas is the so-called Hall current, yielding the Hall MHD (HMHD) model~\citep[see, e.g.][]{T1986}. While this model remains incomplete, at least because it ignores kinetic effects such as the Landau of the cyclotron resonances, it does however provide a useful framework to investigate fundamental features of sub-ion scale plasma dynamics \citep{Bi1997,L1998b,Smi2006,A2009,S2009,A2014a,A2014b}.

Another improvement to the existing exact law models that needs to be achieved is to include density fluctuations $\delta n$. Indeed, while  compressible fluctuations in the solar wind are generally weak ($\delta n/n \sim 10\%-20\%$) and represent only a small fraction of the total MHD fluctuations, which are essentially incompressible and Alfv\'enic~\citep{Kl2014,H2017a,H2017b}, other astrophysical media, e.g. planetary magnetosheaths and the ISM, exhibit higher plasma compressibility ($\delta n/n\sim 50\%-100\%$)~\citep{Sa2006,H2015,H2017b,Z2017}. Moreover, even in the solar wind, the incompressibility assumption can totally fail to describe sub-ion scales physics. This is because Alfv\'en wave turbulence, which is incompressible at MHD scales, transitions into Kinetic Alfv\'enic Wave (KAW) turbulence in the sub-ion scales where density (and pressure) fluctuations become important and couple to the increasing parallel magnetic fluctuations as the energy cascade approaches the ion scales~\citep{Sa2012,P2012,S2012,Ki2013}.  Since KAW turbulence is thought to be the main channel by which energy flows into the sub-ion scales~\citep{B2005,S2009,S2012,P2012,Ch2013}, it is important to develop theoretical models that incorporate density fluctuations as an essential ingredient of turbulence.  Recently a significant step has been achieved by deriving exact law for compressible isothermal MHD turbulence (CMHD)~\citep{B2013}. This model has been applied to {\it in situ } spacecraft data in the fast and slow solar wind and in the terrestrial magnetosheath to investigate the role of density fluctuations in the turbulence dynamics at the MHD scales. In particular, it has been shown that plasma compressibility enhances both the energy cascade rate and the turbulence anisotropy with respect to the incompressible model~\citep{B2016c,H2017a,H2017b}. This provided new clues to explain the longstanding problem of the solar wind heating~\citep{V2007,B2013}. However, the role of density fluctuations in the sub-ion scales remains unexplored because of the lack of similar exact models that cover those small scales.

In the present Letter, using the full three-dimensional (3D) compressible Hall MHD (CHMHD) set of equations, we derive an exact law for fully developed homogeneous isothermal turbulence. The law considers three important aspects of turbulence that should fill existing gaps in the current fluid models of compressible turbulence in magnetized plasmas: density fluctuations,  the Hall current that controls some of the physics at sub-ion scales and spatial anisotropy due to the mean magnetic field. The new exact law derived here provides for the first time a robust means to compute the amount of the total (compressible) energy that sinks from the MHD inertial range into the sub-ion scales where it is eventually dissipated.


{\it Compressible  Hall MHD model.} The 3D CHMHD equations correspond to the momentum equation for the velocity field $\uh$, the induction equation for the magnetic field \textbf{B} and the continuity equation for the scalar density $\rho$. In addition to that, we consider the differential Gauss's law and the divergence-less equation for the current density $\textbf{J}=(c/4\pi)\boldsymbol\nabla\times\textbf{B}$. Alternatively to \textbf{B} and \textbf{J}, we will use the compressible Alfv\'en velocity $\ua\equiv\textbf{B}/\sqrt{4\pi\rho}$ and the compressible electric density $\ja\equiv\textbf{J}/\rho$. Therefore, the CHMHD set of equations can be cast as,
\begin{align}\nonumber
	\partial_t\uh =& -\uh\cdot\boldsymbol\nabla\uh + \ua\cdot\boldsymbol\nabla\ua - \frac{1}{\rho}\boldsymbol\nabla(P+P_M) \\ \label{model:1}
	& - \ua\cdot(\da) + \textbf{D}_k+\textbf{F}_k, \\ \nonumber
\partial_t\ua =& - (\uh-\la\ja)\cdot\boldsymbol\nabla\ua + \ua\cdot\boldsymbol\nabla(\uh-\la\ja) \\ \label{model:2}
	& - \frac{\ua}{2}(\dv-\la\dja) + \textbf{D}_m, \\ \label{model:3}
\partial_t\rho =& -\boldsymbol\nabla\cdot(\rho\uh), \\ \label{model:4}
	\ua\cdot\boldsymbol\nabla\rho =& -2\rho(\da),  \\ \label{model:5}
	\ja\cdot\boldsymbol\nabla\rho =& -\rho(\dja),
\end{align}
where we have defined the dimensionless ion inertial length $\lambda\equiv d_i/L_0$, where $L_0$ is a characteristic length scale, the pressure $P=c_s^2\rho$ for an isothermal plasma with a constant sound speed $c_s$, the magnetic pressure $P_M\equiv\rho u_\text{A}^2/2$, the large-scale kinetic forcing $\textbf{F}_k$ and the kinetic and magnetic dissipative small-scales terms $\textbf{D}_{k,m}$, respectively.


{\it Exact law derivation.} Similarly to CMHD \citep{A2017}, the total energy is one of the ideal invariants of the CHMHD model, since we are considering a two-fluid description with massless electrons \citep{A2014a}. The total energy can be cast as,
\begin{align}
	E(\textbf{x}) &\equiv \frac{\rho}{2}(\uh\cdot\uh+\ua\cdot\ua) + \rho e
\end{align}
where we have introduced the internal compressible energy for an isothermal plasma $e = c_s^2 \log(\rho/\rho_0)$, with $\rho_0$ a reference density value \citep[see, e.g.][]{Ga2016}. On the other hand, the two-point correlation function associated with the total energy is,
\begin{align}
	R_E(\textbf{x},\textbf{x}') &\equiv  \frac{\rho}{2}(\uh\cdot\uh'+\ua\cdot\ua') + \rho e',
\end{align}
where the prime denotes field evaluation at $\textbf{x}'=\textbf{x}+\boldsymbol\ell$ (being ${\boldsymbol\ell}$ the displacement vector) and the angular bracket $\langle\cdot\rangle$ denotes an ensemble average. Under the homogeneity assumption, the correlation functions depends only on the displacement vector ${\boldsymbol\ell}$ \citep{Ba1953}. For the exact law derivation, a dynamical equation for the correlator $\langle R_E+R_E'\rangle$ is essential, since it is for this particular correlator that we can derive an exact law for fully developed homogeneous turbulence \citep{A2017,A2016b}. Using Eqs. \eqref{model:1}-\eqref{model:5} (evaluated both at points $\textbf{x}$ and $\textbf{x}'$) and basic vector algebra properties, it is possible to calculate each term of $\partial_t\langle R_E+R_E'\rangle$ as,
\begin{widetext}
\begin{align} \nonumber
    \partial_t(\rho\uh\cdot\uh') &= - \boldsymbol\nabla\cdot[(\uh\cdot\uh')\rho\uh] + \boldsymbol\nabla\cdot[(\ua\cdot\uh')\rho\ua] - \boldsymbol\nabla'\cdot[(\uh'\cdot\uh)\rho\uh'] + \boldsymbol\nabla'\cdot[(\ua'\cdot\uh)\rho\ua']  \\ \nonumber
    & -\boldsymbol\nabla\cdot(P\uh') - \boldsymbol\nabla\cdot(P_M\uh') - \frac{\rho}{\rho'}\boldsymbol\nabla'\cdot(P'\uh) - \frac{\rho}{\rho'}\boldsymbol\nabla'\cdot(P_M'\uh) + \rho(\uh\cdot\uh')(\dvp) \\ \label{ns}
    & -(\uh'\cdot\ua)\boldsymbol\nabla\cdot(\rho\ua) - \rho(\uh'\cdot\ua)(\da) -\rho(\uh\cdot\ua')(\dap) - \rho(\uh\cdot\ua')(\dap) + d_k + f_k \\ \nonumber
\partial_t(\rho\ua\cdot\ua') &= - \boldsymbol\nabla\cdot[(\ua\cdot\ua')\rho\uh] + \boldsymbol\nabla\cdot[(\uh\cdot\ua')\rho\ua] - \boldsymbol\nabla'\cdot[(\ua'\cdot\ua)\rho\uh'] + \boldsymbol\nabla'\cdot[(\uh'\cdot\ua)\rho\ua'] \\ \nonumber
    & + \la\big\{\boldsymbol\nabla\cdot[(\ua\cdot\ua')\rho\ja] - \boldsymbol\nabla\cdot[(\ja\cdot\ua')\rho\ua] + \boldsymbol\nabla'\cdot[(\ua'\cdot\ua)\rho\ja'] - \boldsymbol\nabla'\cdot[(\ja'\cdot\ua)\rho\ua'] \big\} \\ \nonumber
    & -\frac{\rho}{2}(\ua\cdot\ua')(\dv) - \frac{\rho}{2}(\ua\cdot\ua')(\dvp) + \rho(\ua\cdot\ua')(\dvp) -(\uh\cdot\ua')\boldsymbol\nabla\cdot(\rho\ua) \\ \nonumber
    &- \rho(\uh'\cdot\ua)(\dap) + \la\big\{\frac{\rho}{2}(\ua\cdot\ua')(\dja) + \frac{\rho}{2}(\ua'\cdot\ua)(\djap) - (\ua'\cdot\ua)\boldsymbol\nabla'\cdot(\rho\ja')\big\} \\ \label{in}
    &- \la\big\{(\ja\cdot\ua')\boldsymbol\nabla\cdot(\rho\ua) + (\ja'\cdot\ua)\boldsymbol\nabla'\cdot(\rho\ua')\big\} + d_m, \\ \label{co}
\partial_t(\rho e') &= - \boldsymbol\nabla'\cdot(\rho e'\uh') - \boldsymbol\nabla'\cdot(\rho e'\uh) - \boldsymbol\nabla'\cdot(P\uh')  + \rho e'(\dvp),
\end{align}
\end{widetext}
where we have defined the dissipation and forcing correlation functions as, $d_k = \textbf{D}_k\cdot\uh'+\textbf{D}_k'\cdot\uh$, $f_k = \textbf{F}_k\cdot\uh'+\textbf{F}_k'\cdot\uh$ and $d_m = \textbf{D}_m\cdot\ua'+\textbf{D}_m'\cdot\ua$, respectively. The small-scale contributions to this derivation come from the terms proportional to $\la$.

Assuming homogeneous turbulence \citep{Ba1953,Ga2011}, i.e. $\langle\boldsymbol\nabla'\cdot(~)\rangle=\boldsymbol\nabla_\ell\cdot\langle\rangle$, $\langle\boldsymbol\nabla\cdot(~)\rangle=-\boldsymbol\nabla_\ell\cdot\langle\rangle$ and $\langle \alpha \rangle = \langle\alpha' \rangle$ (with $\alpha$ any scalar function) and using relations \eqref{ns}-\eqref{co}, the dynamical equation for $\langle R_E+R_E'\rangle$ can be cast as (see Supplemental Material),
\begin{widetext}
\begin{align}\nonumber
    \partial_t \langle R_E + R_E'\rangle &= \frac{1}{2}\boldsymbol\nabla_\ell\cdot\bigg\langle [(\delta(\rho\uh)\cdot\delta\uh+\delta(\rho\ua)\cdot\delta\ua + 2\delta e\delta\rho\big]\delta\uh - [\delta(\rho\uh)\cdot\delta\ua+\delta\uh\cdot\delta(\rho\ua)]\delta\ua \\ \nonumber
    &+ 2\la [(\overline{\rho\ja\times\ua})\times\delta\ua-\delta[\ja\times\ua]\times\overline{\rho\ua}] \bigg\rangle + \frac{1}{2}\langle\big(e'+\frac{u_\text{A}}{2}^{'2}\big)\boldsymbol\nabla\cdot(\rho\uh) + (e+\frac{u_\text{A}}{2}^2\big)\boldsymbol\nabla'\cdot(\rho'\uh')\rangle \\ \nonumber
    &+\langle\big(R_E'-\frac{R_B'+R_B}{2}-E'+\frac{P_M'-P'}{2}\big)(\dv)+\big(R_E-\frac{R_B+R_B'}{2}-E+\frac{P_M-P}{2}\big)(\dvp)\rangle \\ \nonumber
    &+\langle\big[R_H-R_H'-\bar{\rho}(\uh'\cdot\ua)+H'+\la\delta\rho\frac{\ja\cdot\ua'}{2}\big](\da)+\big[R_H'-R_H-\bar{\rho}(\uh\cdot\ua')+H-\la\delta\rho\frac{\ja'\cdot\ua}{2}\big](\dap)\rangle \\ \label{exact0}
    & + \frac{\la}{2}\langle(R_B-R_B')(\dja)+(R_B'-R_B)(\djap)\rangle - \frac{1}{2}\langle\beta^{-1'}\boldsymbol\nabla'\cdot(e'\rho\uh) + \beta^{-1}\boldsymbol\nabla\cdot(e\rho'\uh') \rangle + \mathcal{F} + \mathcal{D},
\end{align}
\end{widetext}
where $\beta\equiv u_\text{A}^2/2c_s^2$ and $H(\textbf{x}) \equiv \rho(\uh\cdot\ua)$ is the density-weighted cross-helicity. The two-point correlation functions associated with magnetic energy and the density-weighted cross helicity are $R_B(\textbf{x},\textbf{x}') \equiv \rho\ua\cdot\ua'/2$ and $R_H(\textbf{x},\textbf{x}') \equiv \rho(\uh\cdot\ua'+\ua\cdot\uh')/2$, respectively. We have also introduced the usual increment $\delta\alpha\equiv\alpha'-\alpha$ and the local mean value $\bar{\alpha}\equiv(\alpha'+\alpha)/2$.


{\it Fully developed turbulence.} To obtain the exact law valid in the inertial range, we adopted the usual assumption for fully developed turbulence, where an asymptotic stationary state is expected to be reached \citep{Ga2011,A2016b}. Assuming an infinite (kinetic and magnetic) Reynolds number with a statistical balance between forcing and dissipation, from Eq.~\eqref{exact0} we obtain the exact law for CHMHD turbulence as,
\begin{widetext}
\begin{align}\nonumber
-2\varepsilon &= \frac{1}{2}\boldsymbol\nabla_\ell\cdot\bigg\langle [(\delta(\rho\uh)\cdot\delta\uh+\delta(\rho\ua)\cdot\delta\ua + 2\delta e\delta\rho\big]\delta\uh - [\delta(\rho\uh)\cdot\delta\ua+\delta\uh\cdot\delta(\rho\ua)]\delta\ua \\ \nonumber
    & + 2\la [(\overline{\rho\ja\times\ua})\times\delta\ua-\delta(\ja\times\ua)\times\overline{\rho\ua}] \bigg\rangle + \frac{1}{2}\langle\big(e'+\frac{u_\text{A}}{2}^{'2}\big)\boldsymbol\nabla\cdot(\rho\uh) + (e+\frac{u_\text{A}}{2}^2\big)\boldsymbol\nabla'\cdot(\rho'\uh')\rangle \\ \nonumber
    &+\langle\big(R_E'-\frac{R_B'+R_B}{2}-E'+\frac{P_M'-P'}{2}\big)(\dv)+\big(R_E-\frac{R_B+R_B'}{2}-E+\frac{P_M-P}{2}\big)(\dvp)\rangle \\ \nonumber
    &+\langle\big[R_H-R_H'-\bar{\rho}(\uh'\cdot\ua)+H'+\la\delta\rho\frac{\ja\cdot\ua'}{2}\big](\da)+\big[R_H'-R_H-\bar{\rho}(\uh\cdot\ua')+H-\la\delta\rho\frac{\ja'\cdot\ua}{2}\big](\dap)\rangle \\ \label{exact}
    & + \frac{\la}{2}\langle(R_B-R_B')(\dja)+(R_B'-R_B)(\djap)\rangle - \frac{1}{2}\langle\beta^{-1'}\boldsymbol\nabla'\cdot(e'\rho\uh) + \beta^{-1}\boldsymbol\nabla\cdot(e\rho'\uh') \rangle,
\end{align}
\end{widetext}
where $\varepsilon$ is the energy cascade (or dissipation) rate.  Eq.~\eqref{exact} is the main result of the present Letter. This equation gives an exact relation for fully developed homogeneous CMHD turbulence that is valid in the MHD inertial range and the sub-ion scales. It generalizes previous exact results \citep{Ga2011,B2013,A2017} by including plasma compressibility, spatial anisotropy and the Hall effect. Eq.~\eqref{exact} gives an accurate mathematical means that can be used to estimate the energy cascade rate of turbulence over a broad range of scales in the inertial and sub-ion (dispersive) ranges without the assumption of isotropy.


{\it Discussion.}
The exact law~\eqref{exact} provides a result that should hold as long as the energy injection rate balances the energy dissipation rate in CHMHD turbulence. In other words, Eq.~\eqref{exact} only requires that dissipation terms gets off all the power injected by the forcing terms. In a compact form, expression~\eqref{exact} can be sketched as,
\begin{align}\nonumber
    -2\varepsilon =&~  \frac{1}{2}\boldsymbol\nabla_{\ell}\cdot\big(\textbf{F}^\text{MHD} + \la \textbf{F}^\text{HMHD}\big) \\ \label{compact}
   &~ + (\text{S}^\text{MHD} + \la \text{S}^\text{HMHD}) +  \text{S}^\text{MHD} _\text{H} + \text{M}^\text{MHD} _\beta,
\end{align}
where the terms with the superscript MHD are those present in the exact law for CMHD turbulence \citep{B2013,A2017}, while the terms with the superscript HMHD represent the new small-scale contributions due to the Hall effect. It is worth mentioning that we recover here the four types of terms reported recently in \citet{A2017} for CMHD turbulence (see Supplemental Material for details). The {\it flux} terms $\textbf{F}^\text{MHD}$, which can be written as the local divergence of increments, correspond to the nonlinear cascade of energy across different scales \citep[see, e.g.][]{ORM1997}. The {\it source} terms $\text{S}^\text{MHD}$ are proportional to the global divergence of the fields $\uh$ and $\ua$, and are related to the dilatation (or contraction) of the plasma. The {\it hybrid} terms $\text{S}^\text{MHD}_\text{H}$ can be considered as source- or flux-like terms, while the $\beta$-{\it dependent} terms $\text{M}^\text{MHD}_\beta$ cannot {\it a priori} be transformed into flux or source terms. We emphasize that the $\beta$-dependent terms are a direct consequence of the gradients of the magnetic pressure in the plasma, and thus have no analogs in hydrodynamic (HD) equations \citep[see,][]{A2017}. Finally, the Hall term brings two new small-scale contributions that are related to $\da$ and $\dja$, and a third flux-like term proportional to $\la$ that cannot be written as a function of increments~\citep[see][]{Ga2011,A2016b}. The Hall effect does not give rise to any hybrid or $\beta$-dependent contribution in the exact law~\eqref{exact}.

Several known results can be recovered here as particular limits of Eq.~\eqref{exact}. For spatial scales much larger than the ion inertial length (i.e., $\lambda<<1)$, assuming that kinetic and magnetic fluctuations are of the same order, the terms proportional to $\la$ can be neglected and Eq.~\eqref{exact} reduces to the CMHD exact law previously reported in the literature \citep{A2017,B2013}. Furthermore, in the hydrodynamic limit, i.e., $\ua=0$, we recover the compressible HD exact result for an isothermal plasma turbulence \citep{Ga2011}. \citet{Ga2008} derived the exact law for incompressible HMHD (IHMHD) turbulence, assuming homogeneity and isotropy. Using the velocity, magnetic and electric current fields, his exact result provided a double scaling relation for large and intermediate scales in the inertial range. Taking the incompressibility limit in Eq.~\eqref{compact}, the source, hybrid and $\beta$-dependent terms tend to zero. Furthermore, $\textbf{F}^\text{MHD}$ tends to the well known incompressible Yaglom term \citep{MY1975} and the new small-scale contribution reduces to,
\begin{align}\nonumber
    \boldsymbol\nabla_\ell\cdot\textbf{F}^\text{HMHD} =& -2[\langle\boldsymbol\nabla\cdot(\textbf{J}\times\textbf{B})\times\textbf{B}'\rangle\\ \nonumber
    & +\langle\boldsymbol\nabla'\cdot(\textbf{J}'\times\textbf{B}')\times\textbf{B}\rangle] \\ \label{hall}
    =&~ 4\boldsymbol\nabla_\ell\cdot\langle(\textbf{J}\times\textbf{B})\times\textbf{B}'\rangle,
\end{align}
where we have used $\langle\boldsymbol\nabla\cdot(\textbf{J}\times\textbf{B})\times\textbf{B}'\rangle=\langle\boldsymbol\nabla'\cdot(\textbf{J}'\times\textbf{B}')\times\textbf{B}\rangle$ thanks to the isotropy assumption \citep[see][]{Ga2011}. Expression \eqref{hall} is the Hall contribution to Eq. (52) in \citet{Ga2008} for fully developed IHMHD.


{\it Summary.} In the study of turbulent flows, exact laws provide an essential tool to analyze and understand the nonlinear cascade of energy. The exact law~\eqref{exact} generalized previous exact results, when small-scale effects and compressibility are take into account in the description, and when the isotropy assumption is relaxed. The new exact law~\eqref{exact} can be used to verify, in numerical simulations and spacecraft observations, whether a given range of scales is inertial or dissipative \citep{S2009,A2009,M2011} since the law must hold in the inertial range (far away from the energy injection or dissipative scales). Furthermore, the law can also be used to estimate the energy cascade rate of turbulence over a broad range of scales that span both the inertial and sub-ion (dispersive) ranges without the assumption of isotropy. This work is very timely in particular because of the availability of {\it in situ} spacecraft data from the recently launched multispacecraft NASA/MMS (Magnetospheric MultiScale) mission~\citep{Bu2016}, which provides us with unprecedented high time resolution of the plasma data. The MMS data should allow us to measure all the terms involved in Eq.~\eqref{exact}, including those involving the electric current, with a sufficient time resolution to probe into the sub-ion scales. If the total cascade rate that would be estimated from spacecraft data can be split into two distinct contributions coming from the MHD and sub-ion scales, i.e. $\varepsilon^\text{MHD}$ and $\varepsilon^\text{HMHD}$, respectively, then the difference between the two energy fluxes $\delta  \varepsilon = \varepsilon^\text{MHD}-\varepsilon^\text{HMHD}$ should provide a first estimation of the energy that is dissipated into ion heating regardless of the actual kinetic process involved in the dissipation. The present results and their expected applications are likely to bring new constraints on the actual theoretical models of sub-ion scale compressible turbulence in magnetized plasmas.

\section*{Acknowledgments}

N.A. is supported through an \'Ecole Polytechnique Postdoctoral Fellowship and by LABEX Plas@Par through a grant managed by the Agence Nationale de la Recherche (ANR), as part of the program “Investissements d’Avenir” under the reference ANR-11-IDEX-0004–02. N.A., S.G. and F.S. acknowledge financial support from Programme National Soleil-Terre (PNST).

\bibliographystyle{apsrev4-1}

\begin{thebibliography}{45}%
\makeatletter
\providecommand \@ifxundefined [1]{%
 \@ifx{#1\undefined}
}%
\providecommand \@ifnum [1]{%
 \ifnum #1\expandafter \@firstoftwo
 \else \expandafter \@secondoftwo
 \fi
}%
\providecommand \@ifx [1]{%
 \ifx #1\expandafter \@firstoftwo
 \else \expandafter \@secondoftwo
 \fi
}%
\providecommand \natexlab [1]{#1}%
\providecommand \enquote  [1]{``#1''}%
\providecommand \bibnamefont  [1]{#1}%
\providecommand \bibfnamefont [1]{#1}%
\providecommand \citenamefont [1]{#1}%
\providecommand \href@noop [0]{\@secondoftwo}%
\providecommand \href [0]{\begingroup \@sanitize@url \@href}%
\providecommand \@href[1]{\@@startlink{#1}\@@href}%
\providecommand \@@href[1]{\endgroup#1\@@endlink}%
\providecommand \@sanitize@url [0]{\catcode `\\12\catcode `\$12\catcode
  `\&12\catcode `\#12\catcode `\^12\catcode `\_12\catcode `\%12\relax}%
\providecommand \@@startlink[1]{}%
\providecommand \@@endlink[0]{}%
\providecommand \url  [0]{\begingroup\@sanitize@url \@url }%
\providecommand \@url [1]{\endgroup\@href {#1}{\urlprefix }}%
\providecommand \urlprefix  [0]{URL }%
\providecommand \Eprint [0]{\href }%
\providecommand \doibase [0]{http://dx.doi.org/}%
\providecommand \selectlanguage [0]{\@gobble}%
\providecommand \bibinfo  [0]{\@secondoftwo}%
\providecommand \bibfield  [0]{\@secondoftwo}%
\providecommand \translation [1]{[#1]}%
\providecommand \BibitemOpen [0]{}%
\providecommand \bibitemStop [0]{}%
\providecommand \bibitemNoStop [0]{.\EOS\space}%
\providecommand \EOS [0]{\spacefactor3000\relax}%
\providecommand \BibitemShut  [1]{\csname bibitem#1\endcsname}%
\let\auto@bib@innerbib\@empty
\bibitem [{\citenamefont {Kulsrud}\ and\ \citenamefont {Pearce}(1969)}]{K1969}%
  \BibitemOpen
  \bibfield  {author} {\bibinfo {author} {\bibfnamefont {R.}~\bibnamefont
  {Kulsrud}}\ and\ \bibinfo {author} {\bibfnamefont {W.~P.}\ \bibnamefont
  {Pearce}},\ }\href@noop {} {\bibfield  {journal} {\bibinfo  {journal} {ApJ}\
  }\textbf {\bibinfo {volume} {156}},\ \bibinfo {pages} {445} (\bibinfo {year}
  {1969})}\BibitemShut {NoStop}%
\bibitem [{\citenamefont {{Schekochihin}}\ \emph {et~al.}(2009)\citenamefont
  {{Schekochihin}}, \citenamefont {{Cowley}}, \citenamefont {{Dorland}},
  \citenamefont {{Hammett}}, \citenamefont {{Howes}}, \citenamefont
  {{Quataert}},\ and\ \citenamefont {{Tatsuno}}}]{Sc2009}%
  \BibitemOpen
  \bibfield  {author} {\bibinfo {author} {\bibfnamefont {A.~A.}\ \bibnamefont
  {{Schekochihin}}}, \bibinfo {author} {\bibfnamefont {S.~C.}\ \bibnamefont
  {{Cowley}}}, \bibinfo {author} {\bibfnamefont {W.}~\bibnamefont {{Dorland}}},
  \bibinfo {author} {\bibfnamefont {G.~W.}\ \bibnamefont {{Hammett}}}, \bibinfo
  {author} {\bibfnamefont {G.~G.}\ \bibnamefont {{Howes}}}, \bibinfo {author}
  {\bibfnamefont {E.}~\bibnamefont {{Quataert}}}, \ and\ \bibinfo {author}
  {\bibfnamefont {T.}~\bibnamefont {{Tatsuno}}},\ }\href@noop {} {\bibfield
  {journal} {\bibinfo  {journal} {ApJs}\ }\textbf {\bibinfo {volume} {182}},\
  \bibinfo {pages} {310} (\bibinfo {year} {2009})}\BibitemShut {NoStop}%
\bibitem [{\citenamefont {Bruno}\ and\ \citenamefont {Carbone}(2013)}]{BC2013}%
  \BibitemOpen
  \bibfield  {author} {\bibinfo {author} {\bibfnamefont {R.}~\bibnamefont
  {Bruno}}\ and\ \bibinfo {author} {\bibfnamefont {V.}~\bibnamefont
  {Carbone}},\ }\href@noop {} {\bibfield  {journal} {\bibinfo  {journal}
  {Living Reviews in Solar Physics}\ }\textbf {\bibinfo {volume} {10}},\
  \bibinfo {pages} {2} (\bibinfo {year} {2013})}\BibitemShut {NoStop}%
\bibitem [{\citenamefont {Chandrasekhar}(1951)}]{Ch1951}%
  \BibitemOpen
  \bibfield  {author} {\bibinfo {author} {\bibfnamefont {S.}~\bibnamefont
  {Chandrasekhar}},\ }\href@noop {} {\bibfield  {journal} {\bibinfo  {journal}
  {Proc. R. Soc. London, Ser A}\ }\textbf {\bibinfo {volume} {204}},\ \bibinfo
  {pages} {435} (\bibinfo {year} {1951})}\BibitemShut {NoStop}%
\bibitem [{\citenamefont {Politano}\ and\ \citenamefont
  {Pouquet}(998a)}]{P1998a}%
  \BibitemOpen
  \bibfield  {author} {\bibinfo {author} {\bibfnamefont {H.}~\bibnamefont
  {Politano}}\ and\ \bibinfo {author} {\bibfnamefont {A.}~\bibnamefont
  {Pouquet}},\ }\href@noop {} {\bibfield  {journal} {\bibinfo  {journal} {Phys.
  Rev. E}\ }\textbf {\bibinfo {volume} {57}},\ \bibinfo {pages} {R21} (\bibinfo
  {year} {1998a})}\BibitemShut {NoStop}%
\bibitem [{\citenamefont {Politano}\ and\ \citenamefont
  {Pouquet}(998b)}]{P1998b}%
  \BibitemOpen
  \bibfield  {author} {\bibinfo {author} {\bibfnamefont {H.}~\bibnamefont
  {Politano}}\ and\ \bibinfo {author} {\bibfnamefont {A.}~\bibnamefont
  {Pouquet}},\ }\href@noop {} {\bibfield  {journal} {\bibinfo  {journal}
  {Geophys. Res. Lett.}\ }\textbf {\bibinfo {volume} {25}},\ \bibinfo {pages}
  {273} (\bibinfo {year} {1998b})}\BibitemShut {NoStop}%
\bibitem [{\citenamefont {Matthaeus}\ \emph {et~al.}(1999)\citenamefont
  {Matthaeus}, \citenamefont {Zank}, \citenamefont {Smith},\ and\ \citenamefont
  {Oughton}}]{M1999}%
  \BibitemOpen
  \bibfield  {author} {\bibinfo {author} {\bibfnamefont {W.~H.}\ \bibnamefont
  {Matthaeus}}, \bibinfo {author} {\bibfnamefont {G.~P.}\ \bibnamefont {Zank}},
  \bibinfo {author} {\bibfnamefont {C.~W.}\ \bibnamefont {Smith}}, \ and\
  \bibinfo {author} {\bibfnamefont {S.}~\bibnamefont {Oughton}},\ }\href@noop
  {} {\bibfield  {journal} {\bibinfo  {journal} {Phys. Rev. Lett.}\ }\textbf
  {\bibinfo {volume} {82}},\ \bibinfo {pages} {3444} (\bibinfo {year}
  {1999})}\BibitemShut {NoStop}%
\bibitem [{\citenamefont {Smith}\ \emph {et~al.}(2006)\citenamefont {Smith},
  \citenamefont {Hamilton}, \citenamefont {Vasquez},\ and\ \citenamefont
  {Leamon}}]{Smi2006}%
  \BibitemOpen
  \bibfield  {author} {\bibinfo {author} {\bibfnamefont {C.~W.}\ \bibnamefont
  {Smith}}, \bibinfo {author} {\bibfnamefont {K.}~\bibnamefont {Hamilton}},
  \bibinfo {author} {\bibfnamefont {B.~J.}\ \bibnamefont {Vasquez}}, \ and\
  \bibinfo {author} {\bibfnamefont {R.~J.}\ \bibnamefont {Leamon}},\
  }\href@noop {} {\bibfield  {journal} {\bibinfo  {journal} {The Astrophysical
  Journal Letters}\ }\textbf {\bibinfo {volume} {645}},\ \bibinfo {pages} {L85}
  (\bibinfo {year} {2006})}\BibitemShut {NoStop}%
\bibitem [{\citenamefont {Sorriso-Valvo}\ \emph {et~al.}(2007)\citenamefont
  {Sorriso-Valvo}, \citenamefont {Marino}, \citenamefont {Carbone},
  \citenamefont {Noullez}, \citenamefont {Lepreti}, \citenamefont {Veltri},
  \citenamefont {Bruno}, \citenamefont {Bavassano},\ and\ \citenamefont
  {Pietropaolo}}]{SV2007}%
  \BibitemOpen
  \bibfield  {author} {\bibinfo {author} {\bibfnamefont {L.}~\bibnamefont
  {Sorriso-Valvo}}, \bibinfo {author} {\bibfnamefont {R.}~\bibnamefont
  {Marino}}, \bibinfo {author} {\bibfnamefont {V.}~\bibnamefont {Carbone}},
  \bibinfo {author} {\bibfnamefont {A.}~\bibnamefont {Noullez}}, \bibinfo
  {author} {\bibfnamefont {F.}~\bibnamefont {Lepreti}}, \bibinfo {author}
  {\bibfnamefont {P.}~\bibnamefont {Veltri}}, \bibinfo {author} {\bibfnamefont
  {R.}~\bibnamefont {Bruno}}, \bibinfo {author} {\bibfnamefont
  {B.}~\bibnamefont {Bavassano}}, \ and\ \bibinfo {author} {\bibfnamefont
  {E.}~\bibnamefont {Pietropaolo}},\ }\href@noop {} {\bibfield  {journal}
  {\bibinfo  {journal} {Phys. Rev. Lett.}\ }\textbf {\bibinfo {volume} {99}},\
  \bibinfo {pages} {115001} (\bibinfo {year} {2007})}\BibitemShut {NoStop}%
\bibitem [{\citenamefont {Sahraoui}(2008)}]{Sa2008}%
  \BibitemOpen
  \bibfield  {author} {\bibinfo {author} {\bibfnamefont {F.}~\bibnamefont
  {Sahraoui}},\ }\href@noop {} {\bibfield  {journal} {\bibinfo  {journal}
  {Phys. Rev. E}\ }\textbf {\bibinfo {volume} {78}},\ \bibinfo {pages} {026402}
  (\bibinfo {year} {2008})}\BibitemShut {NoStop}%
\bibitem [{\citenamefont {Wan}\ \emph {et~al.}(2012)\citenamefont {Wan},
  \citenamefont {Oughton}, \citenamefont {Servidio},\ and\ \citenamefont
  {Matthaeus}}]{W2012}%
  \BibitemOpen
  \bibfield  {author} {\bibinfo {author} {\bibfnamefont {M.}~\bibnamefont
  {Wan}}, \bibinfo {author} {\bibfnamefont {S.}~\bibnamefont {Oughton}},
  \bibinfo {author} {\bibfnamefont {S.}~\bibnamefont {Servidio}}, \ and\
  \bibinfo {author} {\bibfnamefont {W.~H.}\ \bibnamefont {Matthaeus}},\
  }\href@noop {} {\bibfield  {journal} {\bibinfo  {journal} {J. Fluid Mech.}\
  }\textbf {\bibinfo {volume} {697}},\ \bibinfo {pages} {296} (\bibinfo {year}
  {2012})}\BibitemShut {NoStop}%
\bibitem [{\citenamefont {Benzi}\ \emph {et~al.}(1993)\citenamefont {Benzi},
  \citenamefont {Ciliberto}, \citenamefont {Tripiccione}, \citenamefont
  {Baudet}, \citenamefont {Massaioli},\ and\ \citenamefont {Succi}}]{R1993}%
  \BibitemOpen
  \bibfield  {author} {\bibinfo {author} {\bibfnamefont {R.}~\bibnamefont
  {Benzi}}, \bibinfo {author} {\bibfnamefont {S.}~\bibnamefont {Ciliberto}},
  \bibinfo {author} {\bibfnamefont {R.}~\bibnamefont {Tripiccione}}, \bibinfo
  {author} {\bibfnamefont {C.}~\bibnamefont {Baudet}}, \bibinfo {author}
  {\bibfnamefont {F.}~\bibnamefont {Massaioli}}, \ and\ \bibinfo {author}
  {\bibfnamefont {S.}~\bibnamefont {Succi}},\ }\href@noop {} {\bibfield
  {journal} {\bibinfo  {journal} {Phys. Rev. E}\ }\textbf {\bibinfo {volume}
  {48}},\ \bibinfo {pages} {R29} (\bibinfo {year} {1993})}\BibitemShut
  {NoStop}%
\bibitem [{\citenamefont {Grossmann}\ \emph {et~al.}(1997)\citenamefont
  {Grossmann}, \citenamefont {Lohse},\ and\ \citenamefont {Reeh}}]{G1997}%
  \BibitemOpen
  \bibfield  {author} {\bibinfo {author} {\bibfnamefont {S.}~\bibnamefont
  {Grossmann}}, \bibinfo {author} {\bibfnamefont {D.}~\bibnamefont {Lohse}}, \
  and\ \bibinfo {author} {\bibfnamefont {A.}~\bibnamefont {Reeh}},\ }\href@noop
  {} {\bibfield  {journal} {\bibinfo  {journal} {Phys. Rev. E}\ }\textbf
  {\bibinfo {volume} {56}},\ \bibinfo {pages} {5473} (\bibinfo {year}
  {1997})}\BibitemShut {NoStop}%
\bibitem [{\citenamefont {Turner}(1983)}]{T1986}%
  \BibitemOpen
  \bibfield  {author} {\bibinfo {author} {\bibfnamefont {L.}~\bibnamefont
  {Turner}},\ }\href@noop {} {\bibfield  {journal} {\bibinfo  {journal} {IEEE
  Trans. Plasma Sci.}\ }\textbf {\bibinfo {volume} {14}},\ \bibinfo {pages}
  {849} (\bibinfo {year} {1983})}\BibitemShut {NoStop}%
\bibitem [{\citenamefont {Biskamp}\ \emph {et~al.}(1997)\citenamefont
  {Biskamp}, \citenamefont {Schwarz},\ and\ \citenamefont {Drake}}]{Bi1997}%
  \BibitemOpen
  \bibfield  {author} {\bibinfo {author} {\bibfnamefont {D.}~\bibnamefont
  {Biskamp}}, \bibinfo {author} {\bibfnamefont {E.}~\bibnamefont {Schwarz}}, \
  and\ \bibinfo {author} {\bibfnamefont {J.~F.}\ \bibnamefont {Drake}},\
  }\href@noop {} {\bibfield  {journal} {\bibinfo  {journal} {Physics of
  Plasmas}\ }\textbf {\bibinfo {volume} {4}},\ \bibinfo {pages} {1002}
  (\bibinfo {year} {1997})}\BibitemShut {NoStop}%
\bibitem [{\citenamefont {Leamon}\ \emph {et~al.}(1998)\citenamefont {Leamon},
  \citenamefont {Smith}, \citenamefont {Ness}, \citenamefont {Matthaeus},\ and\
  \citenamefont {Wong}}]{L1998b}%
  \BibitemOpen
  \bibfield  {author} {\bibinfo {author} {\bibfnamefont {R.~J.}\ \bibnamefont
  {Leamon}}, \bibinfo {author} {\bibfnamefont {C.~W.}\ \bibnamefont {Smith}},
  \bibinfo {author} {\bibfnamefont {N.~F.}\ \bibnamefont {Ness}}, \bibinfo
  {author} {\bibfnamefont {W.~H.}\ \bibnamefont {Matthaeus}}, \ and\ \bibinfo
  {author} {\bibfnamefont {H.~K.}\ \bibnamefont {Wong}},\ }\href {\doibase
  10.1029/97JA03394} {\bibfield  {journal} {\bibinfo  {journal} {J. Geophys. Res.}\ }\textbf {\bibinfo
  {volume} {103}},\ \bibinfo {pages} {4775} (\bibinfo {year}
  {1998})}\BibitemShut {NoStop}%
\bibitem [{\citenamefont {Alexandrova}\ \emph {et~al.}(2009)\citenamefont
  {Alexandrova}, \citenamefont {Saur}, \citenamefont {Lacombe}, \citenamefont
  {Mangeney}, \citenamefont {Mitchell}, \citenamefont {Schwartz},\ and\
  \citenamefont {Robert}}]{A2009}%
  \BibitemOpen
  \bibfield  {author} {\bibinfo {author} {\bibfnamefont {O.}~\bibnamefont
  {Alexandrova}}, \bibinfo {author} {\bibfnamefont {J.}~\bibnamefont {Saur}},
  \bibinfo {author} {\bibfnamefont {C.}~\bibnamefont {Lacombe}}, \bibinfo
  {author} {\bibfnamefont {A.}~\bibnamefont {Mangeney}}, \bibinfo {author}
  {\bibfnamefont {J.}~\bibnamefont {Mitchell}}, \bibinfo {author}
  {\bibfnamefont {S.~J.}\ \bibnamefont {Schwartz}}, \ and\ \bibinfo {author}
  {\bibfnamefont {P.}~\bibnamefont {Robert}},\ }\href@noop {} {\bibfield
  {journal} {\bibinfo  {journal} {Phys. Rev. Lett.}\ }\textbf {\bibinfo
  {volume} {103}},\ \bibinfo {pages} {165003} (\bibinfo {year}
  {2009})}\BibitemShut {NoStop}%
\bibitem [{\citenamefont {Sahraoui}\ \emph {et~al.}(2009)\citenamefont
  {Sahraoui}, \citenamefont {Goldstein}, \citenamefont {Robert},\ and\
  \citenamefont {Khotyaintsev}}]{S2009}%
  \BibitemOpen
  \bibfield  {author} {\bibinfo {author} {\bibfnamefont {F.}~\bibnamefont
  {Sahraoui}}, \bibinfo {author} {\bibfnamefont {M.~L.}\ \bibnamefont
  {Goldstein}}, \bibinfo {author} {\bibfnamefont {P.}~\bibnamefont {Robert}}, \
  and\ \bibinfo {author} {\bibfnamefont {Y.~V.}\ \bibnamefont {Khotyaintsev}},\
  }\href@noop {} {\bibfield  {journal} {\bibinfo  {journal} {Phys. Rev. Lett.}\
  }\textbf {\bibinfo {volume} {102}},\ \bibinfo {pages} {231102} (\bibinfo
  {year} {2009})}\BibitemShut {NoStop}%
\bibitem [{\citenamefont {Andr\'es}\ \emph
  {et~al.}(2014{\natexlab{a}})\citenamefont {Andr\'es}, \citenamefont {Martin},
  \citenamefont {Dmitruk},\ and\ \citenamefont {G\'omez}}]{A2014a}%
  \BibitemOpen
  \bibfield  {author} {\bibinfo {author} {\bibfnamefont {N.}~\bibnamefont
  {Andr\'es}}, \bibinfo {author} {\bibfnamefont {L.~N.}\ \bibnamefont
  {Martin}}, \bibinfo {author} {\bibfnamefont {P.}~\bibnamefont {Dmitruk}}, \
  and\ \bibinfo {author} {\bibfnamefont {D.~O.}\ \bibnamefont {G\'omez}},\
  }\href@noop {} {\bibfield  {journal} {\bibinfo  {journal} {Phys. Plasmas}\
  }\textbf {\bibinfo {volume} {21}},\ \bibinfo {pages} {072904} (\bibinfo
  {year} {2014}{\natexlab{a}})}\BibitemShut {NoStop}%
\bibitem [{\citenamefont {Andr\'es}\ \emph
  {et~al.}(2014{\natexlab{b}})\citenamefont {Andr\'es}, \citenamefont
  {Gonzalez}, \citenamefont {Martin}, \citenamefont {Dmitruk},\ and\
  \citenamefont {G\'omez}}]{A2014b}%
  \BibitemOpen
  \bibfield  {author} {\bibinfo {author} {\bibfnamefont {N.}~\bibnamefont
  {Andr\'es}}, \bibinfo {author} {\bibfnamefont {C.}~\bibnamefont {Gonzalez}},
  \bibinfo {author} {\bibfnamefont {L.~N.}\ \bibnamefont {Martin}}, \bibinfo
  {author} {\bibfnamefont {P.}~\bibnamefont {Dmitruk}}, \ and\ \bibinfo
  {author} {\bibfnamefont {D.~O.}\ \bibnamefont {G\'omez}},\ }\href@noop {}
  {\bibfield  {journal} {\bibinfo  {journal} {Phys. Plasmas}\ }\textbf
  {\bibinfo {volume} {21}},\ \bibinfo {pages} {122305} (\bibinfo {year}
  {2014}{\natexlab{b}})}\BibitemShut {NoStop}%
\bibitem [{\citenamefont {Klein}\ \emph {et~al.}(2014)\citenamefont {Klein},
  \citenamefont {Howes}, \citenamefont {TenBarge},\ and\ \citenamefont
  {Podesta}}]{Kl2014}%
  \BibitemOpen
  \bibfield  {author} {\bibinfo {author} {\bibfnamefont {K.~G.}\ \bibnamefont
  {Klein}}, \bibinfo {author} {\bibfnamefont {G.~G.}\ \bibnamefont {Howes}},
  \bibinfo {author} {\bibfnamefont {J.~M.}\ \bibnamefont {TenBarge}}, \ and\
  \bibinfo {author} {\bibfnamefont {J.~J.}\ \bibnamefont {Podesta}},\
  }\href@noop {} {\bibfield  {journal} {\bibinfo  {journal} {The Astrophysical
  Journal}\ }\textbf {\bibinfo {volume} {785}},\ \bibinfo {pages} {138}
  (\bibinfo {year} {2014})}\BibitemShut {NoStop}%
\bibitem [{\citenamefont {Hadid}\ \emph {et~al.}(017a)\citenamefont {Hadid},
  \citenamefont {Sahraoui},\ and\ \citenamefont {Galtier}}]{H2017a}%
  \BibitemOpen
  \bibfield  {author} {\bibinfo {author} {\bibfnamefont {L.~Z.}\ \bibnamefont
  {Hadid}}, \bibinfo {author} {\bibfnamefont {F.}~\bibnamefont {Sahraoui}}, \
  and\ \bibinfo {author} {\bibfnamefont {S.}~\bibnamefont {Galtier}},\
  }\href@noop {} {\bibfield  {journal} {\bibinfo  {journal} {ApJ}\ }\textbf
  {\bibinfo {volume} {9}},\ \bibinfo {pages} {838} (\bibinfo {year}
  {2017a})}\BibitemShut {NoStop}%
\bibitem [{\citenamefont {Hadid}\ \emph {et~al.}(017b)\citenamefont {Hadid},
  \citenamefont {Sahraoui},\ and\ \citenamefont {Galtier}}]{H2017b}%
  \BibitemOpen
  \bibfield  {author} {\bibinfo {author} {\bibfnamefont {L.~Z.}\ \bibnamefont
  {Hadid}}, \bibinfo {author} {\bibfnamefont {F.}~\bibnamefont {Sahraoui}}, \
  and\ \bibinfo {author} {\bibfnamefont {S.}~\bibnamefont {Galtier}},\
  }\href@noop {} {\bibfield  {journal} {\bibinfo  {journal} {Manuscript
  submitted}\ } (\bibinfo {year} {2017b})}\BibitemShut {NoStop}%
\bibitem [{\citenamefont {Sahraoui}\ \emph {et~al.}(2006)\citenamefont
  {Sahraoui}, \citenamefont {Belmont}, \citenamefont {Rezeau}, \citenamefont
  {Cornilleau-Wehrlin}, \citenamefont {Pin{\c{c}}on},\ and\ \citenamefont
  {Balogh}}]{Sa2006}%
  \BibitemOpen
  \bibfield  {author} {\bibinfo {author} {\bibfnamefont {F.}~\bibnamefont
  {Sahraoui}}, \bibinfo {author} {\bibfnamefont {G.}~\bibnamefont {Belmont}},
  \bibinfo {author} {\bibfnamefont {L.}~\bibnamefont {Rezeau}}, \bibinfo
  {author} {\bibfnamefont {N.}~\bibnamefont {Cornilleau-Wehrlin}}, \bibinfo
  {author} {\bibfnamefont {J.}~\bibnamefont {Pin{\c{c}}on}}, \ and\ \bibinfo
  {author} {\bibfnamefont {A.}~\bibnamefont {Balogh}},\ }\href@noop {}
  {\bibfield  {journal} {\bibinfo  {journal} {Physical review letters}\
  }\textbf {\bibinfo {volume} {96}},\ \bibinfo {pages} {075002} (\bibinfo
  {year} {2006})}\BibitemShut {NoStop}%
\bibitem [{\citenamefont {Hadid}\ \emph {et~al.}(2015)\citenamefont {Hadid},
  \citenamefont {Sahraoui}, \citenamefont {Kiyani}, \citenamefont {Retinò},
  \citenamefont {Modolo}, \citenamefont {Canu}, \citenamefont {Masters},\ and\
  \citenamefont {Dougherty}}]{H2015}%
  \BibitemOpen
  \bibfield  {author} {\bibinfo {author} {\bibfnamefont {L.~Z.}\ \bibnamefont
  {Hadid}}, \bibinfo {author} {\bibfnamefont {F.}~\bibnamefont {Sahraoui}},
  \bibinfo {author} {\bibfnamefont {K.~H.}\ \bibnamefont {Kiyani}}, \bibinfo
  {author} {\bibfnamefont {A.}~\bibnamefont {Retinò}}, \bibinfo {author}
  {\bibfnamefont {R.}~\bibnamefont {Modolo}}, \bibinfo {author} {\bibfnamefont
  {P.}~\bibnamefont {Canu}}, \bibinfo {author} {\bibfnamefont {A.}~\bibnamefont
  {Masters}}, \ and\ \bibinfo {author} {\bibfnamefont {M.~K.}\ \bibnamefont
  {Dougherty}},\ }\href@noop {} {\bibfield  {journal} {\bibinfo  {journal} {The
  Astrophysical Journal Letters}\ }\textbf {\bibinfo {volume} {813}},\ \bibinfo
  {pages} {L29} (\bibinfo {year} {2015})}\BibitemShut {NoStop}%
\bibitem [{\citenamefont {Zank}\ \emph {et~al.}(2017)\citenamefont {Zank},
  \citenamefont {Du},\ and\ \citenamefont {Hunana}}]{Z2017}%
  \BibitemOpen
  \bibfield  {author} {\bibinfo {author} {\bibfnamefont {G.}~\bibnamefont
  {Zank}}, \bibinfo {author} {\bibfnamefont {S.}~\bibnamefont {Du}}, \ and\
  \bibinfo {author} {\bibfnamefont {P.}~\bibnamefont {Hunana}},\ }\href@noop {}
  {\bibfield  {journal} {\bibinfo  {journal} {The Astrophysical Journal}\
  }\textbf {\bibinfo {volume} {842}},\ \bibinfo {pages} {16pp} (\bibinfo {year}
  {2017})}\BibitemShut {NoStop}%
\bibitem [{\citenamefont {Sahraoui}\ \emph {et~al.}(2012)\citenamefont
  {Sahraoui}, \citenamefont {Belmont},\ and\ \citenamefont
  {Goldstein}}]{Sa2012}%
  \BibitemOpen
  \bibfield  {author} {\bibinfo {author} {\bibfnamefont {F.}~\bibnamefont
  {Sahraoui}}, \bibinfo {author} {\bibfnamefont {G.}~\bibnamefont {Belmont}}, \
  and\ \bibinfo {author} {\bibfnamefont {M.~L.}\ \bibnamefont {Goldstein}},\
  }\href@noop {} {\bibfield  {journal} {\bibinfo  {journal} {The Astrophysical
  Journal}\ }\textbf {\bibinfo {volume} {748}},\ \bibinfo {pages} {100}
  (\bibinfo {year} {2012})}\BibitemShut {NoStop}%
\bibitem [{\citenamefont {Podesta}\ and\ \citenamefont
  {TenBarge}(2012)}]{P2012}%
  \BibitemOpen
  \bibfield  {author} {\bibinfo {author} {\bibfnamefont {J.~J.}\ \bibnamefont
  {Podesta}}\ and\ \bibinfo {author} {\bibfnamefont {J.~M.}\ \bibnamefont
  {TenBarge}},\ }\href@noop {} {\bibfield  {journal} {\bibinfo  {journal}
  {Journal of Geophysical Research: Space Physics}\ }\textbf {\bibinfo {volume}
  {117}} (\bibinfo {year} {2012})}\BibitemShut {NoStop}%
\bibitem [{\citenamefont {Salem}\ \emph {et~al.}(2012)\citenamefont {Salem},
  \citenamefont {Howes}, \citenamefont {Sundkvist}, \citenamefont {Bale},
  \citenamefont {Chaston}, \citenamefont {Chen},\ and\ \citenamefont
  {Mozer}}]{S2012}%
  \BibitemOpen
  \bibfield  {author} {\bibinfo {author} {\bibfnamefont {C.}~\bibnamefont
  {Salem}}, \bibinfo {author} {\bibfnamefont {G.}~\bibnamefont {Howes}},
  \bibinfo {author} {\bibfnamefont {D.}~\bibnamefont {Sundkvist}}, \bibinfo
  {author} {\bibfnamefont {S.}~\bibnamefont {Bale}}, \bibinfo {author}
  {\bibfnamefont {C.}~\bibnamefont {Chaston}}, \bibinfo {author} {\bibfnamefont
  {C.}~\bibnamefont {Chen}}, \ and\ \bibinfo {author} {\bibfnamefont
  {F.}~\bibnamefont {Mozer}},\ }\href@noop {} {\bibfield  {journal} {\bibinfo
  {journal} {The Astrophysical Journal Letters}\ }\textbf {\bibinfo {volume}
  {745}},\ \bibinfo {pages} {L9} (\bibinfo {year} {2012})}\BibitemShut
  {NoStop}%
\bibitem [{\citenamefont {Kiyani}\ \emph {et~al.}(2013)\citenamefont {Kiyani},
  \citenamefont {Chapman}, \citenamefont {Sahraoui}, \citenamefont {Hnat},
  \citenamefont {Fauvarque},\ and\ \citenamefont {Khotyaintsev}}]{Ki2013}%
  \BibitemOpen
  \bibfield  {author} {\bibinfo {author} {\bibfnamefont {K.~H.}\ \bibnamefont
  {Kiyani}}, \bibinfo {author} {\bibfnamefont {S.~C.}\ \bibnamefont {Chapman}},
  \bibinfo {author} {\bibfnamefont {F.}~\bibnamefont {Sahraoui}}, \bibinfo
  {author} {\bibfnamefont {B.}~\bibnamefont {Hnat}}, \bibinfo {author}
  {\bibfnamefont {O.}~\bibnamefont {Fauvarque}}, \ and\ \bibinfo {author}
  {\bibfnamefont {Y.~V.}\ \bibnamefont {Khotyaintsev}},\ }\href
  {http://stacks.iop.org/0004-637X/763/i=1/a=10} {\bibfield  {journal}
  {\bibinfo  {journal} {The Astrophysical Journal}\ }\textbf {\bibinfo {volume}
  {763}},\ \bibinfo {pages} {10} (\bibinfo {year} {2013})}\BibitemShut
  {NoStop}%
\bibitem [{\citenamefont {Bale}\ \emph {et~al.}(2005)\citenamefont {Bale},
  \citenamefont {Kellogg}, \citenamefont {Mozer}, \citenamefont {Horbury},\
  and\ \citenamefont {Reme}}]{B2005}%
  \BibitemOpen
  \bibfield  {author} {\bibinfo {author} {\bibfnamefont {S.}~\bibnamefont
  {Bale}}, \bibinfo {author} {\bibfnamefont {P.}~\bibnamefont {Kellogg}},
  \bibinfo {author} {\bibfnamefont {F.}~\bibnamefont {Mozer}}, \bibinfo
  {author} {\bibfnamefont {T.}~\bibnamefont {Horbury}}, \ and\ \bibinfo
  {author} {\bibfnamefont {H.}~\bibnamefont {Reme}},\ }\href@noop {} {\bibfield
   {journal} {\bibinfo  {journal} {Physical Review Letters}\ }\textbf {\bibinfo
  {volume} {94}},\ \bibinfo {pages} {215002} (\bibinfo {year}
  {2005})}\BibitemShut {NoStop}%
\bibitem [{\citenamefont {Chen}\ \emph {et~al.}(2013)\citenamefont {Chen},
  \citenamefont {Boldyrev}, \citenamefont {Xia},\ and\ \citenamefont
  {Perez}}]{Ch2013}%
  \BibitemOpen
  \bibfield  {author} {\bibinfo {author} {\bibfnamefont {C.}~\bibnamefont
  {Chen}}, \bibinfo {author} {\bibfnamefont {S.}~\bibnamefont {Boldyrev}},
  \bibinfo {author} {\bibfnamefont {Q.}~\bibnamefont {Xia}}, \ and\ \bibinfo
  {author} {\bibfnamefont {J.}~\bibnamefont {Perez}},\ }\href@noop {}
  {\bibfield  {journal} {\bibinfo  {journal} {Physical review letters}\
  }\textbf {\bibinfo {volume} {110}},\ \bibinfo {pages} {225002} (\bibinfo
  {year} {2013})}\BibitemShut {NoStop}%
\bibitem [{\citenamefont {Banerjee}\ and\ \citenamefont
  {Galtier}(2013)}]{B2013}%
  \BibitemOpen
  \bibfield  {author} {\bibinfo {author} {\bibfnamefont {S.}~\bibnamefont
  {Banerjee}}\ and\ \bibinfo {author} {\bibfnamefont {S.}~\bibnamefont
  {Galtier}},\ }\href@noop {} {\bibfield  {journal} {\bibinfo  {journal} {Phys.
  Rev. E}\ }\textbf {\bibinfo {volume} {87}},\ \bibinfo {pages} {013019}
  (\bibinfo {year} {2013})}\BibitemShut {NoStop}%
\bibitem [{\citenamefont {Banerjee}\ \emph {et~al.}(2016)\citenamefont
  {Banerjee}, \citenamefont {Hadid}, \citenamefont {Sahraoui},\ and\
  \citenamefont {Galtier}}]{B2016c}%
  \BibitemOpen
  \bibfield  {author} {\bibinfo {author} {\bibfnamefont {S.}~\bibnamefont
  {Banerjee}}, \bibinfo {author} {\bibfnamefont {L.~Z.}\ \bibnamefont {Hadid}},
  \bibinfo {author} {\bibfnamefont {F.}~\bibnamefont {Sahraoui}}, \ and\
  \bibinfo {author} {\bibfnamefont {S.}~\bibnamefont {Galtier}},\ }\href@noop
  {} {\bibfield  {journal} {\bibinfo  {journal} {ApJs}\ }\textbf {\bibinfo
  {volume} {829}},\ \bibinfo {pages} {L27} (\bibinfo {year}
  {2016})}\BibitemShut {NoStop}%
\bibitem [{\citenamefont {Vasquez}\ \emph {et~al.}(2007)\citenamefont
  {Vasquez}, \citenamefont {Smith}, \citenamefont {Hamilton}, \citenamefont
  {MacBride},\ and\ \citenamefont {Leamon}}]{V2007}%
  \BibitemOpen
  \bibfield  {author} {\bibinfo {author} {\bibfnamefont {B.~J.}\ \bibnamefont
  {Vasquez}}, \bibinfo {author} {\bibfnamefont {C.~W.}\ \bibnamefont {Smith}},
  \bibinfo {author} {\bibfnamefont {K.}~\bibnamefont {Hamilton}}, \bibinfo
  {author} {\bibfnamefont {B.~T.}\ \bibnamefont {MacBride}}, \ and\ \bibinfo
  {author} {\bibfnamefont {R.~J.}\ \bibnamefont {Leamon}},\ }\href@noop {}
  {\bibfield  {journal} {\bibinfo  {journal} {Journal of Geophysical Research:
  Space Physics}\ }\textbf {\bibinfo {volume} {112}} (\bibinfo {year}
  {2007})}\BibitemShut {NoStop}%
\bibitem [{\citenamefont {Andr\'es}\ and\ \citenamefont
  {Sahraoui}(2017)}]{A2017}%
  \BibitemOpen
  \bibfield  {author} {\bibinfo {author} {\bibfnamefont {N.}~\bibnamefont
  {Andr\'es}}\ and\ \bibinfo {author} {\bibfnamefont {F.}~\bibnamefont
  {Sahraoui}},\ }\href@noop {} {\bibfield  {journal} {\bibinfo  {journal}
  {Manuscript submitted, arXiv:1707.00749}\ } (\bibinfo {year}
  {2017})}\BibitemShut {NoStop}%
\bibitem [{\citenamefont {Galtier}(2016)}]{Ga2016}%
  \BibitemOpen
  \bibfield  {author} {\bibinfo {author} {\bibfnamefont {S.}~\bibnamefont
  {Galtier}},\ }\href@noop {} {\emph {\bibinfo {title} {Introduction to modern
  magnetohydrodynamics}}}\ (\bibinfo  {publisher} {Cambridge University
  Press},\ \bibinfo {year} {2016})\BibitemShut {NoStop}%
\bibitem [{\citenamefont {Batchelor}(1953)}]{Ba1953}%
  \BibitemOpen
  \bibfield  {author} {\bibinfo {author} {\bibfnamefont {G.~K.}\ \bibnamefont
  {Batchelor}},\ }\href@noop {} {\emph {\bibinfo {title} {The theory of
  homogeneus turbulence}}}\ (\bibinfo  {publisher} {Cambridge Univ. Press},\
  \bibinfo {year} {1953})\BibitemShut {NoStop}%
\bibitem [{\citenamefont {Andr\'es}\ \emph {et~al.}(016b)\citenamefont
  {Andr\'es}, \citenamefont {Mininni}, \citenamefont {Dmitruk},\ and\
  \citenamefont {G\'omez}}]{A2016b}%
  \BibitemOpen
  \bibfield  {author} {\bibinfo {author} {\bibfnamefont {N.}~\bibnamefont
  {Andr\'es}}, \bibinfo {author} {\bibfnamefont {P.}~\bibnamefont {Mininni}},
  \bibinfo {author} {\bibfnamefont {P.}~\bibnamefont {Dmitruk}}, \ and\
  \bibinfo {author} {\bibfnamefont {D.~O.}\ \bibnamefont {G\'omez}},\
  }\href@noop {} {\bibfield  {journal} {\bibinfo  {journal} {Phys. Rev. E}\
  }\textbf {\bibinfo {volume} {93}},\ \bibinfo {pages} {063202} (\bibinfo
  {year} {2016b})}\BibitemShut {NoStop}%
\bibitem [{\citenamefont {Galtier}\ and\ \citenamefont
  {Banerjee}(2011)}]{Ga2011}%
  \BibitemOpen
  \bibfield  {author} {\bibinfo {author} {\bibfnamefont {S.}~\bibnamefont
  {Galtier}}\ and\ \bibinfo {author} {\bibfnamefont {S.}~\bibnamefont
  {Banerjee}},\ }\href@noop {} {\bibfield  {journal} {\bibinfo  {journal}
  {Phys. Rev. Lett.}\ }\textbf {\bibinfo {volume} {107}},\ \bibinfo {pages}
  {134501} (\bibinfo {year} {2011})}\BibitemShut {NoStop}%
\bibitem [{\citenamefont {Oughton}\ \emph {et~al.}(1997)\citenamefont
  {Oughton}, \citenamefont {Radler},\ and\ \citenamefont
  {Matthaeus}}]{ORM1997}%
  \BibitemOpen
  \bibfield  {author} {\bibinfo {author} {\bibfnamefont {S.}~\bibnamefont
  {Oughton}}, \bibinfo {author} {\bibfnamefont {K.~H.}\ \bibnamefont {Radler}},
  \ and\ \bibinfo {author} {\bibfnamefont {W.~H.}\ \bibnamefont {Matthaeus}},\
  }\href@noop {} {\bibfield  {journal} {\bibinfo  {journal} {Phys. Rev. E}\
  }\textbf {\bibinfo {volume} {56}},\ \bibinfo {pages} {2875} (\bibinfo {year}
  {1997})}\BibitemShut {NoStop}%
\bibitem [{\citenamefont {Galtier}(2008)}]{Ga2008}%
  \BibitemOpen
  \bibfield  {author} {\bibinfo {author} {\bibfnamefont {S.}~\bibnamefont
  {Galtier}},\ }\href@noop {} {\bibfield  {journal} {\bibinfo  {journal} {Phys.
  Rev. E}\ }\textbf {\bibinfo {volume} {77}},\ \bibinfo {pages} {015302}
  (\bibinfo {year} {2008})}\BibitemShut {NoStop}%
\bibitem [{\citenamefont {Monin}\ and\ \citenamefont {Yaglom}(1975)}]{MY1975}%
  \BibitemOpen
  \bibfield  {author} {\bibinfo {author} {\bibfnamefont {A.~S.}\ \bibnamefont
  {Monin}}\ and\ \bibinfo {author} {\bibfnamefont {A.~M.}\ \bibnamefont
  {Yaglom}},\ }\href@noop {} {\emph {\bibinfo {title} {Statistical Fluid
  Mechanics: Mechanics of Turbulence}}},\ Vol.~\bibinfo {volume} {2}\ (\bibinfo
   {publisher} {Cambridge, MA: MIT Press.},\ \bibinfo {year}
  {1975})\BibitemShut {NoStop}%
\bibitem [{\citenamefont {Matthaeus}\ and\ \citenamefont
  {Velli}(2011)}]{M2011}%
  \BibitemOpen
  \bibfield  {author} {\bibinfo {author} {\bibfnamefont {W.}~\bibnamefont
  {Matthaeus}}\ and\ \bibinfo {author} {\bibfnamefont {M.}~\bibnamefont
  {Velli}},\ }\href@noop {} {\bibfield  {journal} {\bibinfo  {journal} {Space
  Science Reviews}\ }\textbf {\bibinfo {volume} {160}},\ \bibinfo {pages} {145}
  (\bibinfo {year} {2011})}\BibitemShut {NoStop}%
\bibitem [{\citenamefont {Burch}\ \emph {et~al.}(2016)\citenamefont {Burch},
  \citenamefont {Moore}, \citenamefont {Torbert},\ and\ \citenamefont
  {Giles}}]{Bu2016}%
  \BibitemOpen
  \bibfield  {author} {\bibinfo {author} {\bibfnamefont {J.~L.}\ \bibnamefont
  {Burch}}, \bibinfo {author} {\bibfnamefont {T.~E.}\ \bibnamefont {Moore}},
  \bibinfo {author} {\bibfnamefont {R.~B.}\ \bibnamefont {Torbert}}, \ and\
  \bibinfo {author} {\bibfnamefont {B.~L.}\ \bibnamefont {Giles}},\ }\href@noop
  {} {\bibfield  {journal} {\bibinfo  {journal} {Space Science Reviews}\
  }\textbf {\bibinfo {volume} {199}},\ \bibinfo {pages} {5} (\bibinfo {year}
  {2016})}\BibitemShut {NoStop}%
\end{thebibliography}

%

\end{document}